\documentclass[11pt]{article}

\usepackage[final]{acl}
\usepackage{hyperref}
\usepackage{times}
\usepackage{latexsym}
\usepackage{placeins}
\usepackage[T1]{fontenc}

\usepackage[utf8]{inputenc}

\usepackage{microtype}

\usepackage{inconsolata}

\usepackage{graphicx}

\title{Modeling Changing Scientific Concepts with Complex Networks: A Case Study on the Chemical Revolution}

\author{Sofía Aguilar-Valdez \and Stefania Degaetano-Ortlieb
\\
Language Science and Technology Department, 
Saarland University 
\\
\small{\texttt{\href{mailto:sofia.aguilar@uni-saarland.de}{sofia.aguilar@uni-saarland.de}, \href{mailto:s.degaetano@mx.uni-saarland.de}{s.degaetano@mx.uni-saarland.de}}}
}

\begin{document}
\maketitle
\begin{abstract}
While context embeddings produced by LLMs can be used to estimate conceptual change, these representations are often not interpretable nor time-aware. Moreover, bias augmentation in historical data poses a non-trivial risk to researchers in the Digital Humanities. Hence, to model reliable concept trajectories in evolving scholarship, in this work we develop a framework that represents prototypical concepts through complex networks based on topics. Utilizing the Royal Society Corpus, we analyzed two competing theories from the Chemical Revolution (phlogiston vs.\ oxygen) as a case study to show that onomasiological change is linked to higher entropy and topological density, indicating increased diversity of ideas and connectivity effort.
\end{abstract}

\section{Introduction}
Language change is not random; it is driven by shifting communicative goals, social structures, and domain-specific conventions \citep{Hamilton_2016_DiachronicWordEmbeddings, Gries_2008_IdentificationStagesDiachronica, Blank_2013_WhyNewMeanings}. While various methods exist to model change in language use, such as lexical semantic change through word meaning representations from LLMs, interpretability remains a difficult task \citep{Periti_2024_LexicalSemanticChange}. This impedes domain experts' ability to qualitatively assess change and adapt models to their specific needs \citep{Beck_2024_ReviewTahmasebiBorin}. 

We model conceptual change through language use in the context of scientific revolutions, where shifts in language use reflect broader epistemic transitions. Specifically, we want to analyze how the lexemes referring to a concept evolve after discoveries \cite{Zgusta_2011_OnomasiologicalChangeSachenchange}, which poses the challenges of quantifying concepts with adaptable structures and predicting what conceptual evolution mechanisms are associated with epistemic drifts. To this end, we propose a graph-based approach that models scientific texts as concept networks, organizing a concept's lexemes or readings from core to periphery, and evaluates the recombination of these readings into new core concepts for onomasiological\footnote{Onomasiological change refers to shifts in the linguistic expressions used to denote the same underlying concept. Instead of tracking how a word’s meaning evolves (a semasiological perspective), the onomasiological approach fixes a concept and studies competition, replacement, or specialization among its lexical realizations over time (cf.\ \citet{Geeraerts2010})} assessment.


\begin{figure}[t]
  \includegraphics[width=\columnwidth]{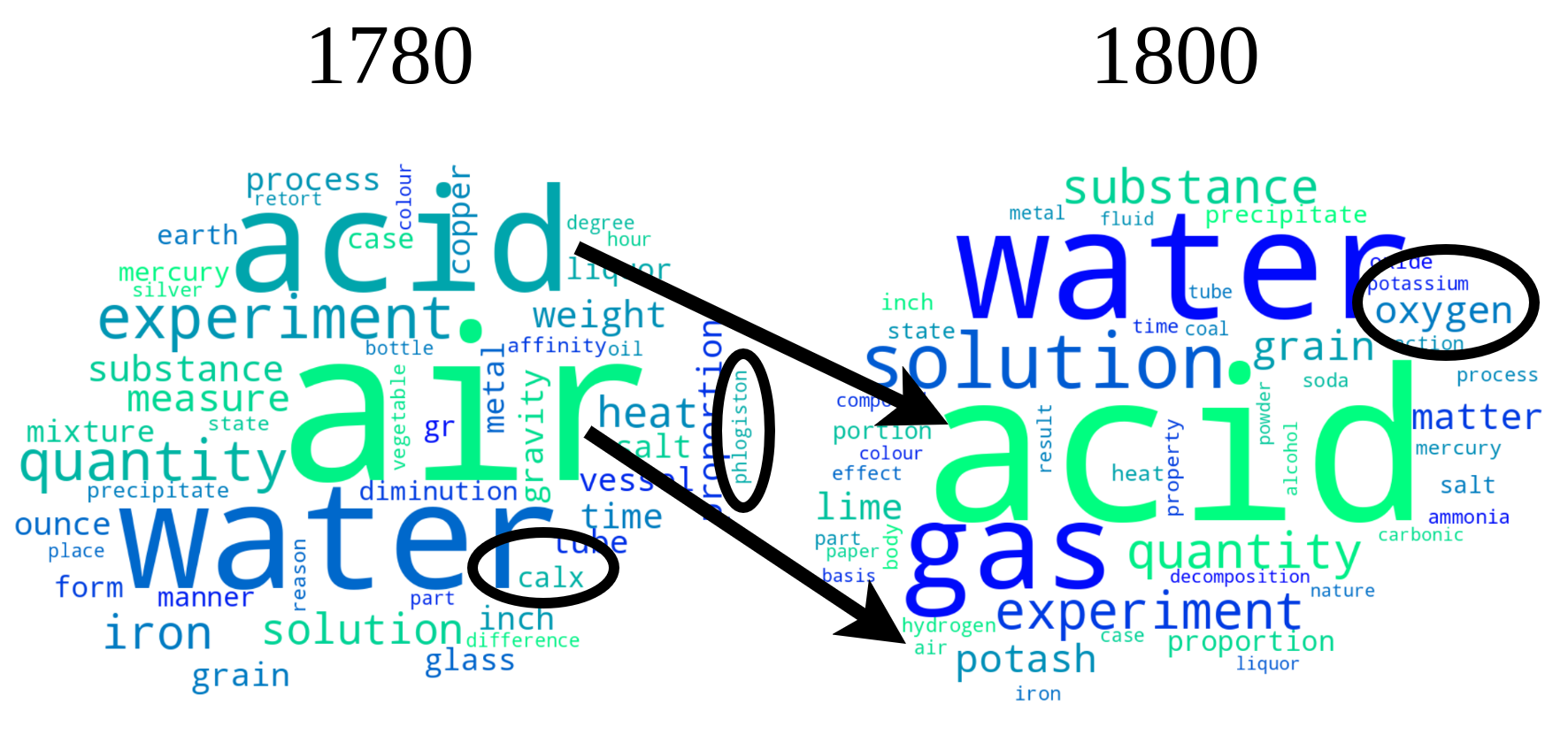}
  \caption {\textbf{Diachronic prototypical concepts}. This schematic illustrates movements in central and peripheral readings (arrows), and the presence of notable terms (ovals). Over time, \textit{air} went from the core to the periphery while \textit{acid} did the opposite, and this was accompanied by the removal of \textit{calx},   \textit{phlogiston}, and the emergence of \textit{oxygen}.}
  \label{schema}
\end{figure}

As a case study, we analyze two competing explanations for combustion during the Chemical Revolution: phlogiston theory (e.g., “a lighter burns because it contains phlogiston”) and oxygen theory (e.g., “a lighter burns because some substances ignite when in contact with oxygen”). Considering these theories as prototypical concepts (see Figure \ref{schema}) with central and peripheral readings \citep{Geeraerts_1997_DiachronicPrototypeSemantics}, the overarching research question of this work is concerned with modeling onomasiological change in scientific discourse with temporal graphs. Our central question is whether the network structure reflects the spread and competition of readings in changing scientific concepts. 

The rest of the paper is organized as follows. Section 2 motivates the study of concepts, especially scientific ones, and explains how prototype semantics serves as the theoretical framework for representing concepts as networks. Section 3 reviews related work on modeling concept formation using networks and introduces our task of detecting conceptual structure shifts. Section 4 describes the corpus and experiments, and Section 5 reports the results and their interpretation. Finally, Section 6 concludes and provides future research directions.

\section{Background}
This section begins by explaining why modeling concepts is important, especially in the scientific domain, and presents the linguistic theory from which we approach concepts as graphs. It ends with a brief historical background on the Chemical Revolution.

\subsection{Concepts and Networks}
\paragraph{Why Concepts Matter} Concepts are mental tools designed to manage the (im)balance between expectation and experience \citep{Koselleck_2006_VerzeitlichungBegriffe}. For instance, the concept of \textit{nature} today is approached as an object rather than subject, which presents interpretations that diverge from those of previous societies. Since concepts are indicators of historical change, tracing their shifts is key to reconstruct how past societies understood and structured their world.

\paragraph{Why Scientific Concepts} Unlike everyday concepts (e.g., \textit{chair}), scientific concepts (e.g., \textit{DNA}) cannot be formed by the subject through immediate experience of the object. They are learned through the academic establishment and lose meaning if isolated from this universal system of related ideas since they rely on neither moral judgments, gut feelings nor authority. This is how \citet{Vygotsky_1994_DevelopmentThinkingConcepta} explains why scientific concepts are "true concepts" (viz., truly learned, nonspontaneous concepts). Since they are not impacted by cultural ambiguity, as  study objects they enable "universal sensors" for historical change.

\paragraph{Why Complex Networks} According to prototype semantics, concepts have readings or interpretations located at different levels; going from the center to the periphery, where the centroids are considered to be the most typical \citep{Geeraerts_1997_DiachronicPrototypeSemantics}. To illustrate this, consider how the concept of \textit{fruit} generates a typical reading that varies depending on the cultural perspective. This structure is akin to a graph's, and the literature supports that representing them as such offers means to model the meaning dynamics as well as social interactions that lead to concept formation \citep{Kaye_2024_SocioEpistemicNetworksFramework, Kedrick_2024_ConceptualStructureGrowth, Ju_2020_NetworkStructureScientific}.

\subsection{Chemical Revolution Overview} 
When science undergoes dramatic ontological changes (i.e., the whole system of concepts and their laws is replaced by a new one), by analogy with political uprisings, such processes are called scientific revolutions \citep{Kuhn_2012_StructureScientificRevolutions}. The textbook example of this occurred in the 18th century, when competing sides disagreed sharply about what counted as an element and what was a compound, a foundational difference many chemists of the time had \citep{Chang_2015_ChemicalRevolutionRevisited}. Notably, Lavoisier's oxygen paradigm overtook the dominant one in chemistry for combustion, Stahl's phlogiston paradigm; while phlogistonists regarded metals as calx compounds (i.e., today known as oxides) plus phlogiston, oxygenists classified calxes as compounds of pure metals plus oxygen (i.e., today known as element), which cleared out research questions at the time related to not only combustion but calcination and respiration \citep{Thagard_1990_ConceptualStructureChemical}, producing the ground work for many additional transformations in chemistry \citep{Holmes_1994_WhatWasChemical}. Arguably, phlogistonists' contributions were hindered by interpreting phenomena using phlogiston as a principle or "central reading", an ontological category that disappeared as Lavoisier's system developed \citep{Chang_2015_ChemicalRevolutionRevisited}.

 Hence, this scientific scenario is optimal as a case study to model how concepts are changed, new ones arise, and old ones disappear. In particular, we focus on the decades where the new concept of oxygen became established while phlogiston vanished: from the 1750s to 1800s \citep{Chang_2011_PersistenceEpistemicObjects, Thagard_1990_ConceptualStructureChemical}. This scenario, however, only serves as a first step towards a generalizable approach to model conceptual transitions and the meaning dynamics involved in these changes.

\section{Related Work}
This section introduces the state-of-the-art methods for the NLP task most related to our problem (i.e., lexical-semantic change detection), the current challenge of improving interpretability, and how we approach it in this work as a conceptual structure shift detection task.  

\subsection{NLP Research Overview} The evolution of word meanings, often termed semantic change, has long been a central focus of linguistic inquiry. This process is crucial for understanding how language adapts to shifting cultural and communicative contexts. Advances in computational methods have provided novel tools to study semantic change at scale. \citet{Tahmasebi-etal2021} and \citet{Periti_2024_LexicalSemanticChange} offer comprehensive surveys on these methods, with a focus on diachronic word embeddings and neural methods. Diachronic embeddings capture the shifting meanings of words across time by mapping their usage in successive temporal intervals \citep{Hamilton_2016_DiachronicWordEmbeddings,dubossarsky-etal-2017-outta}. While such approaches effectively track semantic drift, they often fall short in allowing to understand the underlying mechanisms driving these changes.

Modeling evolving scholarship considering concept networks offers an alternative, more informative perspective. By building semantic graphs that organize a concept's core and peripheral readings, \citet{Kedrick_2024_ConceptualStructureGrowth} found conceptual structures with rigid cores hinder scientific innovation. They explain this using the network structure: the recombination of cores with fellow cores and peripheral entities is less likely to occur in disciplines where core concepts are more stable. Such framing is compatible with prototype semantics, which describes words as radial structures (e.g., word \textit{fruit} with \textit{apple} and \textit{guava} as central and peripheral readings, respectively) where depending on the situational context some readings are more "central" or "typical" than others, boundaries are not rigid, and different concepts can overlap \citep{Geeraerts_1997_DiachronicPrototypeSemantics}. 

Networks are thus informative in the sense that they can represent the meaning dynamics taking place alongside the social interactions driving them. By constructing community networks based on a topic model, \citet{Malaterre_2023_IdentifyingHiddenCommunities} showed the evolution of communities and their main research themes, tracing the development of the specialization areas that structure the field of philosophy of science today. Topic models are known for approaching documents as mixtures of abstract themes, where each theme is represented by a probability distribution over words \citep{Blei_2003_LatentDirichletAllocation}. Document-topic distributions thus offer insight into how specific themes are distributed across time, and to our knowledge this approach was first implemented by \citet{Griffiths_2004_FindingScientificTopics} to find scientific disciplines and evaluate collaborations between researchers. 

\subsection{Towards a Conceptual Structure Shift Detection Task}
We approach concepts as prototypical structures whose meaning evolves through their relational embedding in domain-specific discourse. The conceptual structure shift detection task therefore frames conceptual change as a network-level phenomenon rather than a purely lexical one.

\begin{itemize}
    \item \textbf{Input}: A diachronic scientific corpus (e.g., spanning multiple decades of publications from one or more research domains).
    \item \textbf{Output}: Concept networks for each time slice that capture core–periphery organization and their diachronic trajectories.
\end{itemize}

In the context of scientific revolutions, these trajectories should reflect conceptual structure shifts that we define as \textit{periods where the prototypical core of a concept either reconfigures or expands towards new peripheries, resulting in multiple meaning shift cases}, namely 
\begin{itemize}
    \item the lexemes referring to the core diversify (i.e., \textbf{onomasiological change}),
    \item new overall senses for the core emerge (i.e., \textbf{semasiological change}),
    \item and/or a new ontological system replaces the previous one (i.e., \textbf{ontological change}).
\end{itemize}
None of these changes are exclusive and arguably happen in sequence (e.g., during the Chemical Revolution, onomasiological/semasiological changes led to the ontological one). Moreover, to explain these cases systems are expected to produce both static models per period and dynamic mappings across periods, \textbf{prioritizing interpretability in structural metrics}.

Hence, to alleviate interpretability concerns we estimate concepts as themes with a ranked vocabulary using topic modeling, then cluster the documents based on similarity. Doing so produces graphs with topic clusters that represent the concepts present in a decade, and how these concepts were structured in terms of core and peripheral words. 

Regarding evaluation, this baseline adopts two established metrics from information theory to measure change:
\begin{itemize}
    \item \textbf{Jensen-Shannon distance}, to quantify the difference across topic clusters by treating graphs as diachronic, referent-independent entities (i.e., a perspective looking from the future into the past and vice versa have a distance that is symmetrical).
    \item \textbf{Entropy}, to describe the magnitude of the change understood as "topic diversity" according to \citet{Hall_2008_StudyingHistoryIdeasa}.
\end{itemize}

In summary, we approach concepts as prototypical structures and define our task as conceptual structure shift detection: the input is a diachronic scientific corpus, and the expected output are concept networks that offer core and peripheral readings trajectories to inform the \textbf{onomasiological change assessment}, which is the case we decided to focus on for this work. 

\section{Material and Methods}

\begin{figure*}[t]
  \includegraphics[width=0.48\linewidth]{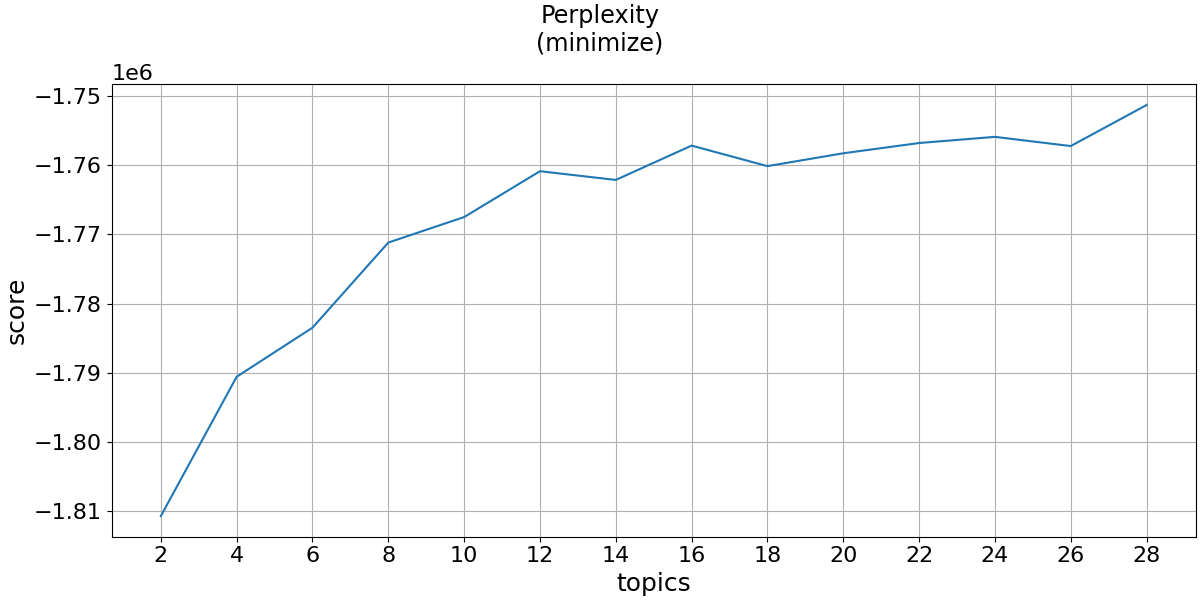} \hfill
  \includegraphics[width=0.48\linewidth]{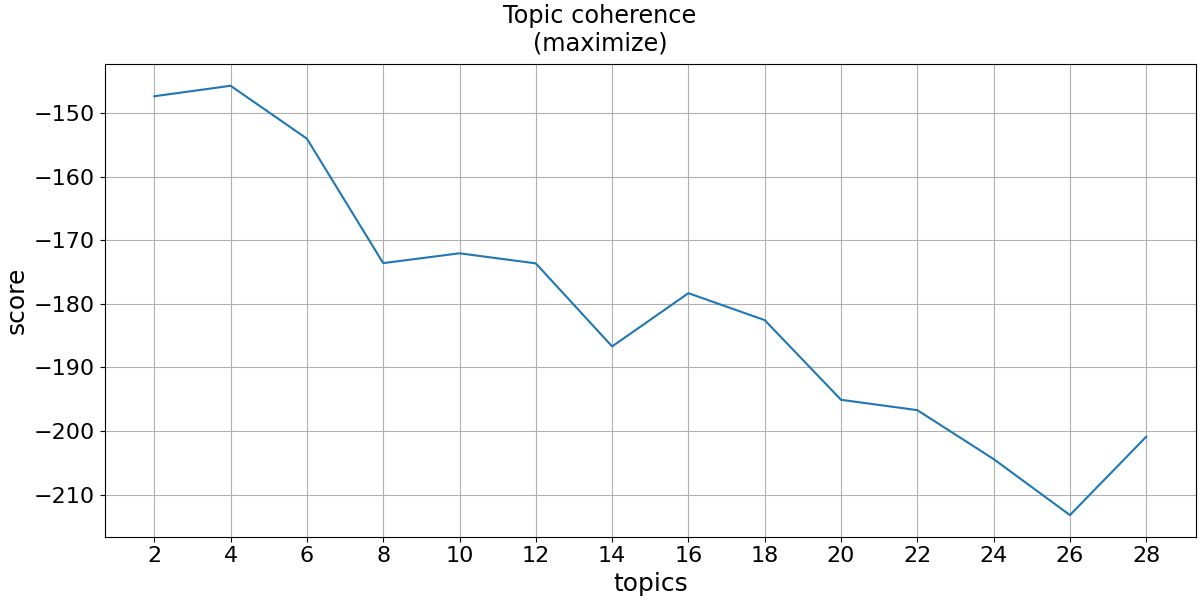}
  \caption {\textbf{Topic models evaluation}. These results, produced by evaluating models for the 1800s non-cumulative corpus, were consistent across decades and strategies. Since both metrics degrade beyond 6 topics (↑perplexity, ↓coherence), the optimal number is 6.}
  \label{eval}
\end{figure*}

Based on the Royal Society Corpus (RSC; \citealt{Kermes_2016_RoyalSocietyCorpus, Fischer_2020_RoyalSocietyCorpus}), this approach consisted of four stages: first, create a representative corpus by filtering publications given specified decades and oxygen terminology \citep{Chang_2011_PersistenceEpistemicObjects, Bizzoni_2021_DiffusionScientificTerms}. Second, represent concepts and their diachronic readings using topic modeling \citep{Blei_2003_LatentDirichletAllocation, sievert2014ldavis}. Third, form concept networks by linking documents based on the Jensen-Shannon distance \citep{shannon2001mathematical} among topics and optimize connectivity by minimizing the percolation threshold \citep{Radicchi_2015_PredictingPercolationThresholds}; this is the tipping point where isolated nodes in a graph abruptly form the largest mutually reachable subgraph, with a low threshold indicating extensive connectivity using minimal resources. Finally, we reveal hidden groups by applying clustering \citep{Lukasova_1979_HierarchicalAgglomerativeClustering} and community detection \citep{Blondel_2008_FastUnfoldingCommunities} algorithms.

\paragraph{Dataset Description} The RSC is a diachronic corpus of English scientific writing covering 47,837 publications and their metadata (e.g., publication year, author) in the Philosophical Transactions and Proceedings of the Royal Society of London from 1665 to 1996 \citep{Fischer_2020_RoyalSocietyCorpus}. We focus on 6 decades (1750-1800) to account for the events where oxygen developed as a concept; from 1774, when Joseph Priestley conformed to the theories of the time and introduced the gas as "dephlogisticated air" \citep{Brown_1997_DephlogisticatedAirRevisited}, till 1789, when Lavoisier presented the gas with a new term and led to the widespread adoption of "oxygen" among chemists \citep{Thagard_1990_ConceptualStructureChemical}. Doing so is supported by previous studies based on the RSC that in the same period report documents and lexico-grammatical changes addressing combustion, gases and chemical reactions, key themes in the phlogiston-to-oxygen conceptual transition \citep{Degaetano-Ortlieb_2019_OptimalCodeCommunication, Bizzoni_2020_LinguisticVariationChange, Teich_2021_LessMoreMore}.


\subsection{Data Processing and Sampling} We retrieved texts from the RSC with a publication date between 1750–1800, defined decade batches to make our results comparable with previous studies on this corpus \citep{Teich_2021_LessMoreMore}, then filtered documents to include instances of terms relevant to the discovery of oxygen. These terms were derived from 
\citet{Bizzoni_2021_DiffusionScientificTerms}, which used Kullback-Leibler Divergence (KLD) to identify distinctive terms during the Chemical Revolution; and \citet{Chang_2011_PersistenceEpistemicObjects}, who explains why the term \textit{oxygen} persisted despite theoretical revisions, arguing that its operational meaning remained stable while phlogiston theory lost operational grounding. \citet{Chang_2011_PersistenceEpistemicObjects} also notes that the theoretical meaning of phlogiston was conceptually close to later notions of energy, and describes phlogiston as a precursor to ideas such as free electrons. 

After filtering the documents ($N=2 337$), we performed two cleaning rounds. First, we removed special characters and terms shorter than two characters, converted the text to lowercase, and removed stopwords based on the standard list imported as \texttt{“ENGLISH\_STOP\_WORDS”} using scikit-learn \citep{sklearn_api}; doing so substantially decreased the corpus size by removing prepositions, pronouns and articles (e.g., "a", "he" and "the") while retaining nouns, which according to \citet{Hamilton_2016_CulturalShiftLinguistic} are more sensitive to linguistic changes. Regarding hyphenated terms, these were unchanged to preserve their semantic unity during topic modeling. Second, we filtered for nouns and adjectives, and lemmatized the corpus using tmtoolkit\footnote{\url{https://tmtoolkit.readthedocs.io/en/latest/}}.

To test the influence of prior textual data on downstream analyses, for the six decades we implemented two sampling strategies: 1) cumulative sampling, where each decade includes the indicated decade and all previous decades' documents, and 2) non-cumulative sampling, where each decade includes only documents from that period.

\subsection{Topic Modeling}
In the context of the RSC, topic modeling has previously been used to estimate scientific disciplines \citep{Fankhauser_2016_TopicalDiversificationTime}. The authors employed Latent Dirichlet Allocation (LDA), a probabilistic model designed to uncover abstract categories given a collection of documents \citep{Blei_2003_LatentDirichletAllocation}, and found 30 distinct disciplines and sub-disciplines to describe the publications. 

Our topic modeling workflow consisted of constructing a document-term matrix from the preprocessed corpus, counting the occurrences of each lemmatized term per document. We evaluated model performance using log-likelihood \citep{Bengio_2003_NeuralProbabilisticLanguage} and topic coherence \citep{Mimno_2011_OptimizingSemanticCoherence}; doing so mitigates the often manual and arbitrary parameter selection in LDA \citep{Schmidt_2012_WordsAloneDismantling, Mohr_2013_IntroductionTopicModels}. Following best practices in interpretability, we generated topic labels using LDAvis \citep{sievert2014ldavis}, which are determined based on relevance scores that are more meaningful compared to word rankings (i.e., if \textit{air} is the top-5 word yet its relevance score is higher compared to other terms, this will be the topic label). This is how we defined the optimal number of topics and the topic labels (i.e., the term that "names" the topic) across decades and strategies. 

\begin{figure*}[t]
  \includegraphics[width=0.48\linewidth]{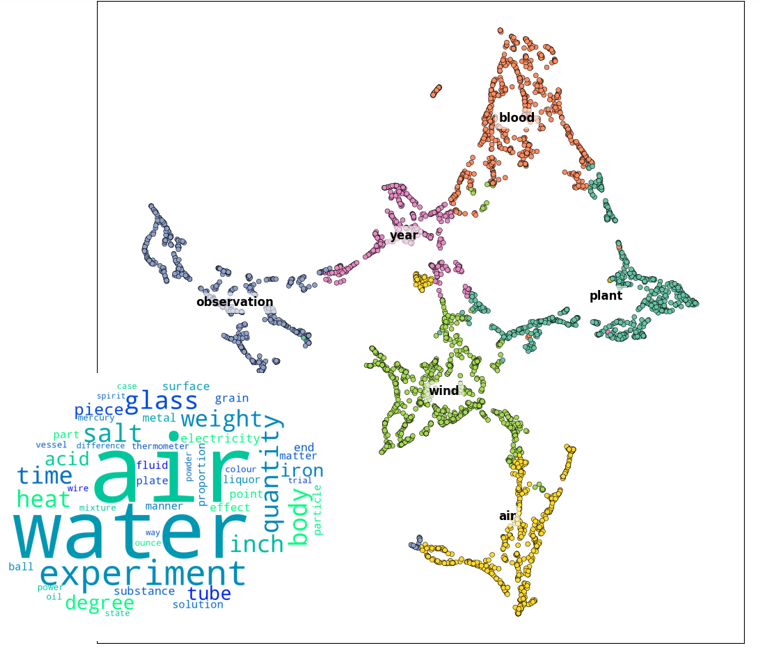} \hfill
  \includegraphics[width=0.48\linewidth]{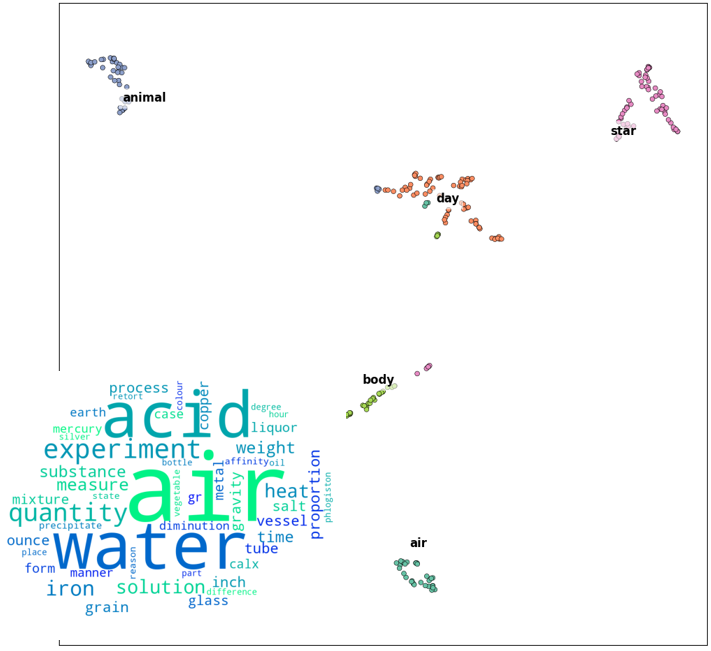} \hfill
  \includegraphics[width=0.48\linewidth]{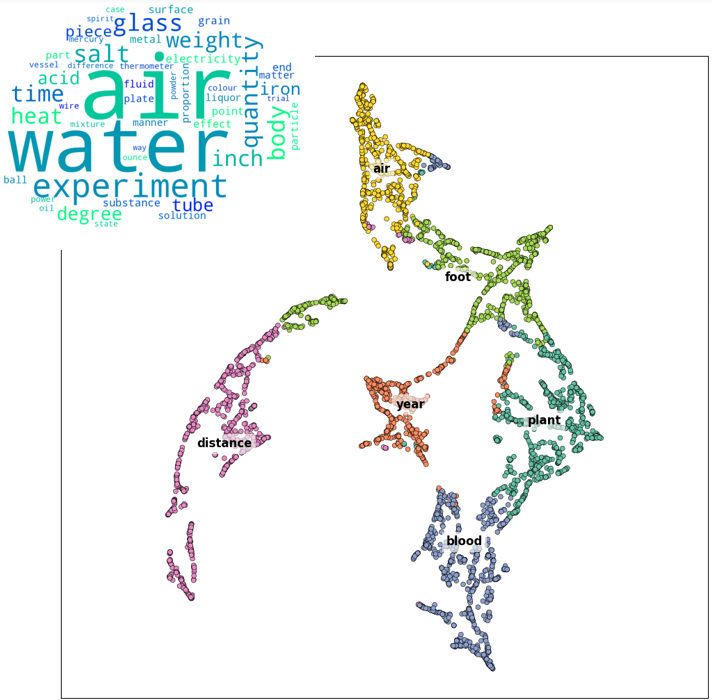} \hfill
  \includegraphics[width=0.48\linewidth]{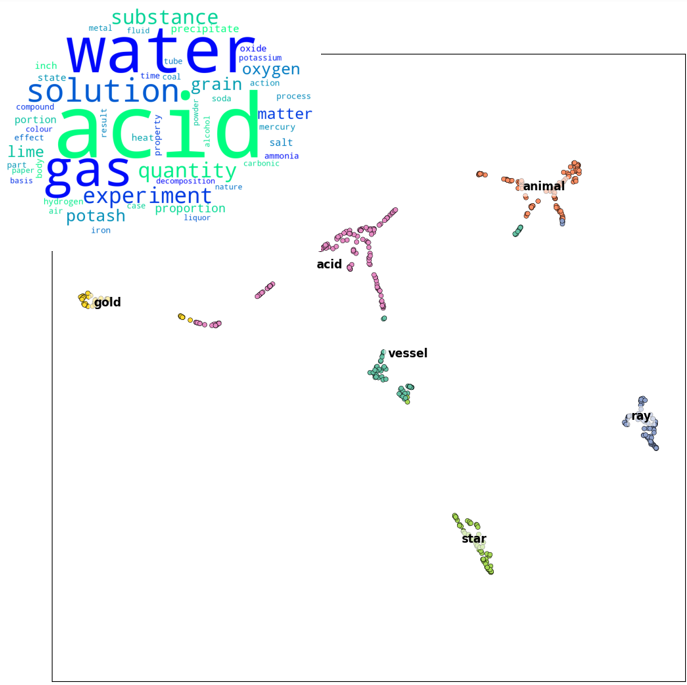} \hfill
  \caption {\textbf{Topic clusters}. By comparing both relevant decades (\texttt{1st row=1780s}, \texttt{2nd row=1800s}) across sampling strategies (\texttt{1st column=cumulative}, \texttt{2nd column=non-cumulative}), we observe the
  cumulative sampling shows topic clusters with stable labels over time, contrary to the non-cumulative that provides more fine-grained representations.}
  \label{topics}
\end{figure*}

\subsection{Network Construction}
We considered documents as nodes and formed edges based on the Jensen-Shannon distance \citep{shannon2001mathematical} between document-topic distributions; given distributions $p$ and $q$, distance $d$ is given by:
$$
d = \sqrt{ \frac{D (p \parallel m) +  D (q \parallel m) }{2}}
$$
where $m$ is the pointwise mean of $p$ and $q$ and $D$ is the KLD.
This way we formed adjacency matrices, later binarized by optimizing connectivity based on the percolation threshold \citep{Radicchi_2015_PredictingPercolationThresholds}. A threshold was necessary to reduce the network's complexity by limiting the number of nodes and edges to only the most meaningful ones.

Hence, the conceptual structure was defined as follows: topic labels \cite{sievert2014ldavis} "name" the concepts, while the topic-word distributions establish the central (i.e., top-10) and peripheral (i.e., top-50) terms.

\subsection{Topic Aggregation}
To find hidden groups, first we applied Hierarchical Agglomerative Clustering (HAC; \citealp{Lukasova_1979_HierarchicalAgglomerativeClustering}), a method that consists of measuring the distance between data points iteratively until forming groups with data points that exhibit higher similarity to one another compared to points in other groups. Each data point starts as its own cluster and merges with others until a number of groups is reached (i.e., 6 according to the optimal number of topics). For visualization purposes, we used cosine similarity as the metric for both HAC and UMAP \citep{mcinnes2018umap} dimensionality reduction approaches due of its reported performance in topic-based networks \citep{Luhmann_2022_DigitalHumanitiesDiscipline}. 

Similarly, the Louvain algorithm \citep{Blondel_2008_FastUnfoldingCommunities} estimates groups by maximizing modularity, a metric that measures the quality of a network’s division into distinct groups or communities, quantifying the density of intra- vs. inter-cluster connections. We used it as a complementary method to further analyze network dynamics among the topics \citep{Malaterre_2023_IdentifyingHiddenCommunities}.

\section{Results and Discussion}
This section describes our findings regarding the phlogiston-to-oxygen conceptual transition: a topic cluster indicates onomasiological change given its recombination of core and peripheral readings (viz., \textit{air} becomes \textit{acid} with \textit{oxygen} as a peripheral term in the non-cumulative strategy), alongside higher entropy, and network metrics capture this shift in terms of rising connectivity effort. We then compare these concept structure dynamics with those reported in qualitative analyses of the Chemical Revolution. 

\subsection{Topic Clusters}
We found that a model of 6 topics seems adequate as it achieved low perplexity and high coherence across all decades (Figure \ref{eval}). 

Figure \ref{topics} shows onomasiological change, where we consider topic clusters as prototype concepts with central (i.e, top-10) and peripheral (i.e., top-50) readings. Notably, with non-cumulative sampling (i.e., no past documents included; 2nd column) in the 1780s corpus (upper graph) at the core is the term \textit{air} taking \textit{acid} as a central reading. This relationship flips by the 1800s (lower graph), when \textit{acid} becomes typical with \textit{oxygen} acting as a peripheral reading (upper right corner of the word cloud). In contrast, cumulative sampling shows topic clusters with recurring labels (e.g., \textit{plant}, \textit{air}).

The observations from non-cumulative sampling are consistent with the conceptual structure reported in the literature that argues \textit{acid} played a pivotal role for the conceptual change \citep{Thagard_1990_ConceptualStructureChemical, Chang_2011_PersistenceEpistemicObjects}: while phlogistonists like Priestley underplayed acidity as a secondary effect of phlogiston release (e.g., experiments burning sulfur produced a "gaseous calx" that was perceived as acidic since it lacked phlogiston), Lavoisier considered acidity as a core explanatory principle to reframe calxes as oxides and oxygen as an enabler of acid formation (e.g., sulfur is a combustible that when in contact with oxygen forms sulfur dioxide, which in turn dissolves in water to produce sulfurous acid). This is explicit in the results, where \textit{calx} and \textit{phlogiston} disappear from the periphery in the concept structure after \textit{acid} takes over the core (cf. notable terms in Figure \ref{schema}).

\begin{figure*}[t]
  \includegraphics[width=0.48\linewidth]{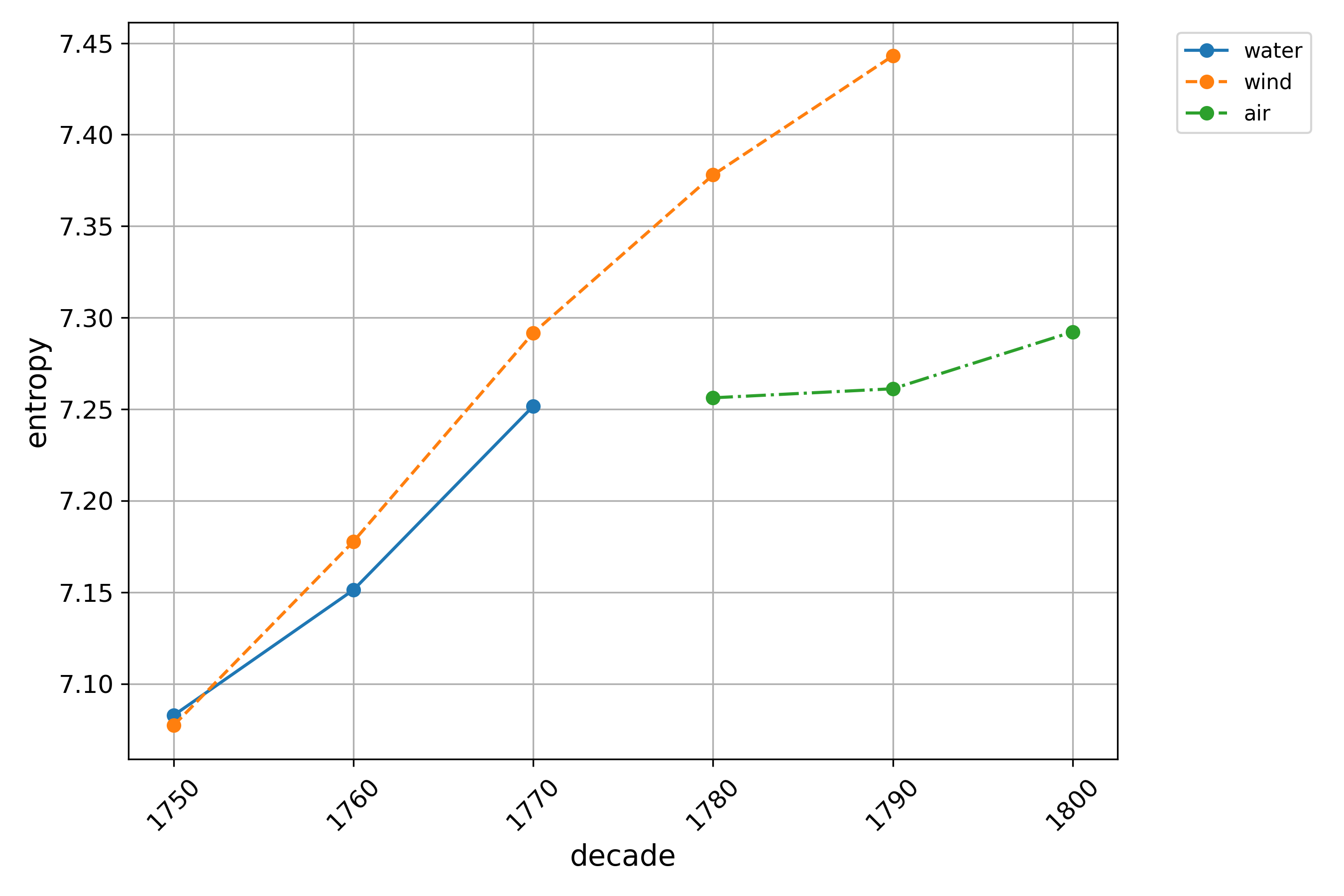} \hfill
  \includegraphics[width=0.48\linewidth]{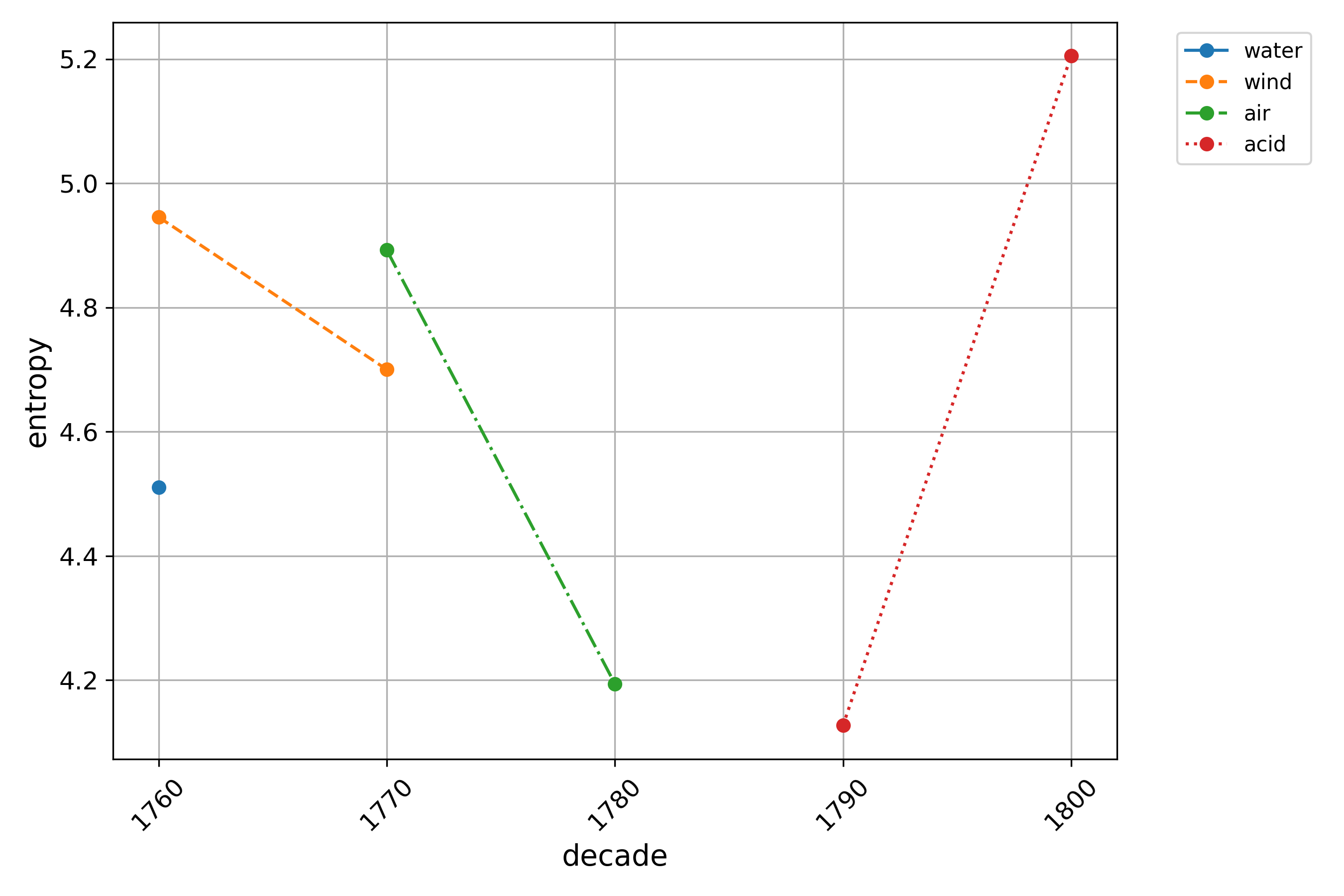}
  \caption {\textbf{Topic entropy over time}. While in cumulative sampling (left) entropy trajectories of oxygen-related terms increase over time, non-cumulative sampling (right) shows a decline then rise during the development of oxygen theory (cf. 1772-1789 in \citet{Thagard_1990_ConceptualStructureChemical}).}
  \label{entropy}
\end{figure*}

To compare the magnitude of diachronic change in both sampling strategies, we measured the topics "diversity of ideas" using entropy \cite{Hall_2008_StudyingHistoryIdeasa}. Figure \ref{entropy} presents different trends: while the cumulative strategy shows rising entropy across oxygen-related topics (illustrated by different colors), the non-cumulative declines then rises post-1774 oxygen discovery. This decline of diversity pre-1780 could be interpreted as the resistance of phlogistonists to de-center \textit{air} in their findings (e.g., Priestley’s "dephlogisticated air"), while the rise of diversity post-1780 could be attributed to Lavoisier coining the term \textit{oxygen}. These observations support \citet{Thagard_1990_ConceptualStructureChemical}'s on why Lavoisier made the conceptual shift instead of Priestley. However, to clarify this, future studies will have to incorporate Lavoisier’s articles in French, which are absent from our corpus. 

\subsection{Temporal Graphs}
Considering the onomasiological change (Figure \ref{topics}) and rising entropy (Figure \ref{entropy}), we built graphs based on the non-cumulative strategy; one for each decade, producing 6 in total (Figure A\ref{app1}). 

To analyze network stability, we considered 5 parameters (Figure A\ref{app2}) from which follow three findings: first, although node size starts and ends similarly (400--500), edge density more than doubled (30 037--62 639). Second, declining network communities suggests integration (5--3), but falling modularity (0,39--0,19) disputes this: fewer, lower-quality communities indicate increased mixing. Lastly, percolation threshold rises in 1760s (0,3--0,54), falls in 1770s (0,54--0,32), then rises (0,32--0,58)—aligning with modularity. 

Low threshold/high modularity signals efficient connectivity given well-defined communities (e.g., communication is faster when people are organized in groups and allotted resources are sufficient), while high threshold/low modularity signals mixed communities requiring high connectivity effort (e.g., mixed groups hinder person-finding, requiring more channels and impeding communication). 


\section{Conclusion and Future Work}
This work case-studies conceptual change during the Chemical Revolution (i.e., phlogiston-to-oxygen) via interpretable complex networks. Our analysis reveals two findings: 1) a topic cluster shifts from \textit{air} to \textit{acid}, showing onomasiological change given the movement of core/peripheral readings accompanied by higher entropy, and 2) network communities and modularity decline, while percolation and edge density rise, indicating higher connectivity effort. These results indicate that changing concepts form high entropy clusters (i.e., more variability of ideas) that increase topological density in the knowledge network (i.e., more effort is needed for community finding and message passing), potentially hindering communication. Future work will refine representation of rare terms and add directed edges (e.g., colexification, citations linked to terms) to account for semantic and social directionality, and predict inter-temporal graph links via node/edge embeddings to identify influential documents and authors.

\section*{Limitations}
Our corpus so far is limited to English scientific texts, which does not account for the whole spectrum of ideas exchanged during the Chemical Revolution. Adding multilingual and multi-genre texts  is a desiderata to work towards a comprehensive picture. In this study, we derived core and peripheral terms from topic-word distributions rather than cluster centroids, potentially overlooking document embedding geometries. Rare peripheral terms receive sparse representations, potentially missing subtle shifts. As an approach that builds undirected graphs, the formed edges overlook semantic directionality; representing this is important to understand how concepts evolve (e.g., through meaning gain or loss), making of the conceptual structure a phylogenetic tree \citep{Carling_2023_EvolutionLexicalSemantics}. Future work will address these challenges via directed graphs and rare-term augmentation, as well as the estimation of phylogenetic structures. Moreover, to test the generalizability of the approach, first steps will include modeling conceptual transitions across disciplines (e.g., germ plasm to chromosomal genes in biology). 

\section*{Acknowledgments} Funded by the European Union under grant 101119511. Views and opinions expressed are however those of the authors only and do not necessarily reflect those of the European Union. Neither the European Union nor the granting authority can be held responsible for them.

\bibliography{custom}

\appendix
\setcounter{figure}{0}
\makeatletter
\renewcommand{\fnum@figure}{Figure A\thefigure}
\makeatother

\begin{figure*}[!t]
  \includegraphics[width=0.48\linewidth]{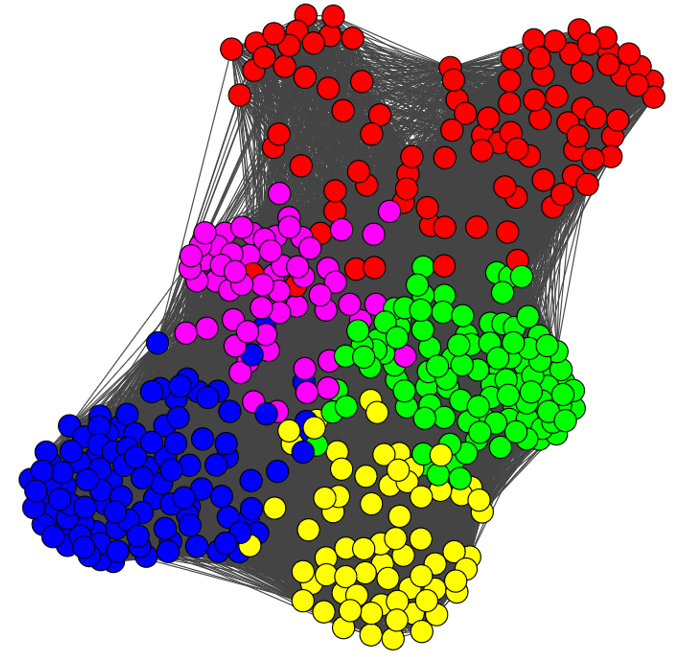} \hfill
  \includegraphics[width=0.48\linewidth]{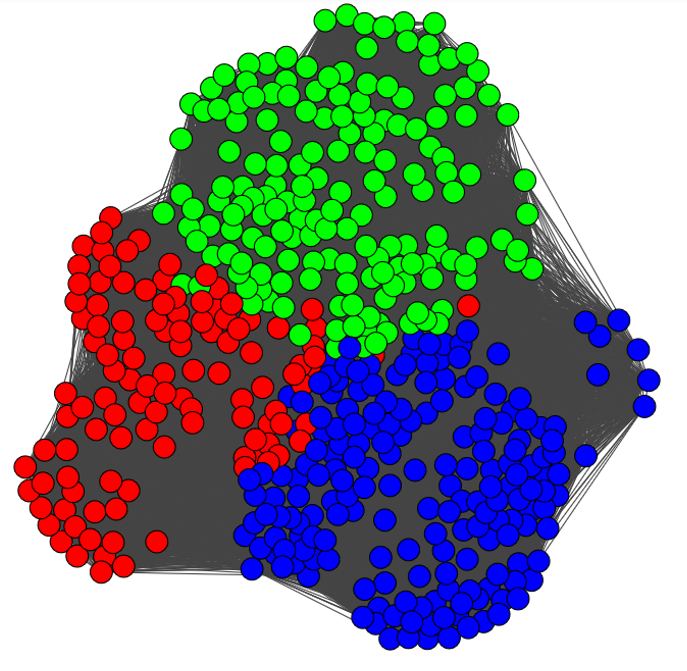} \hfill
  \caption {\textbf{Temporal graphs}. The colors illustrate the number of communities which went on decline: starting at 5 (1750s) and ending at 3 (1800s).}
  \label{app1}
\end{figure*}

\begin{figure*}[!t]
  \includegraphics[width=0.48\linewidth]{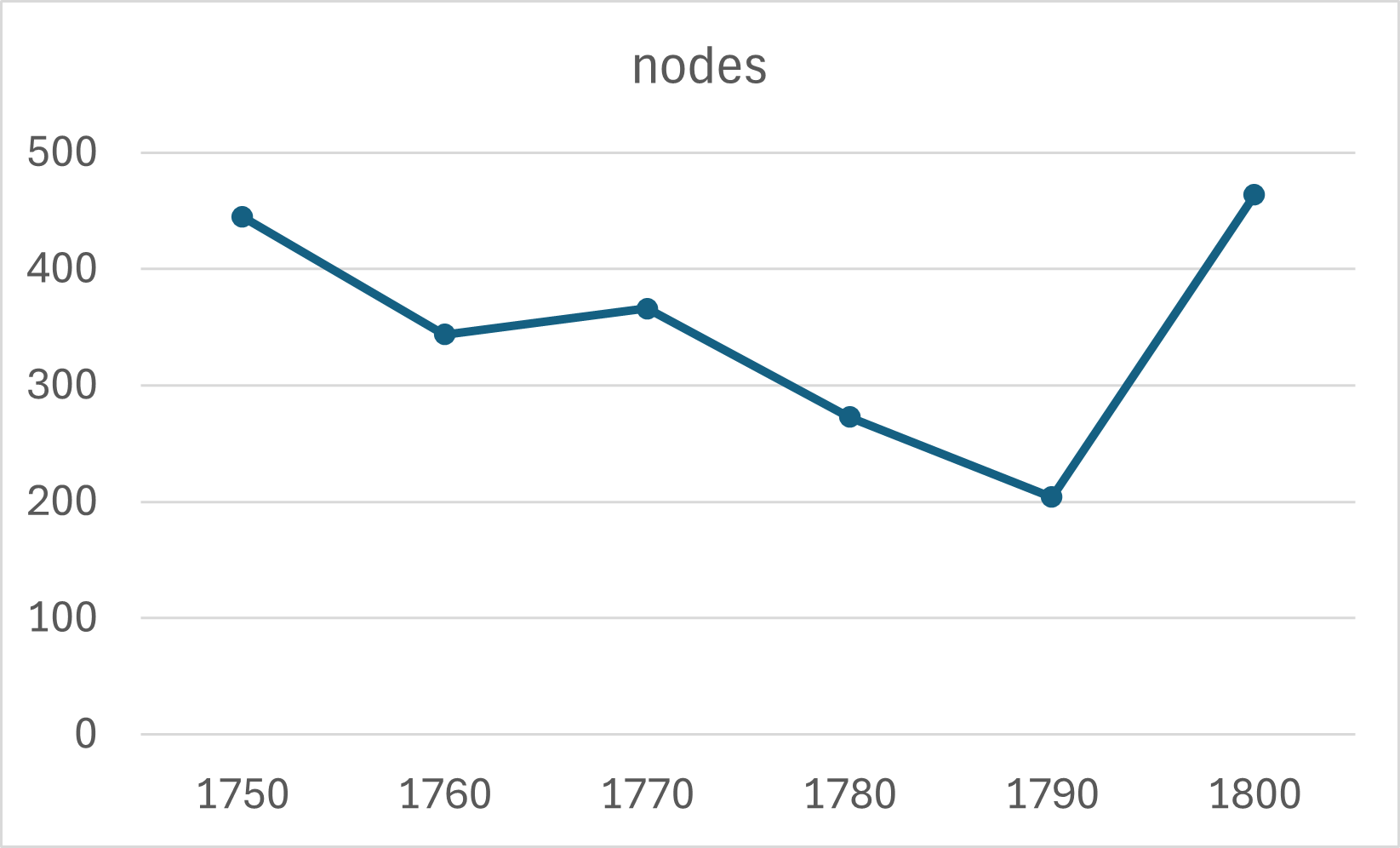} \hfill
  \includegraphics[width=0.48\linewidth]{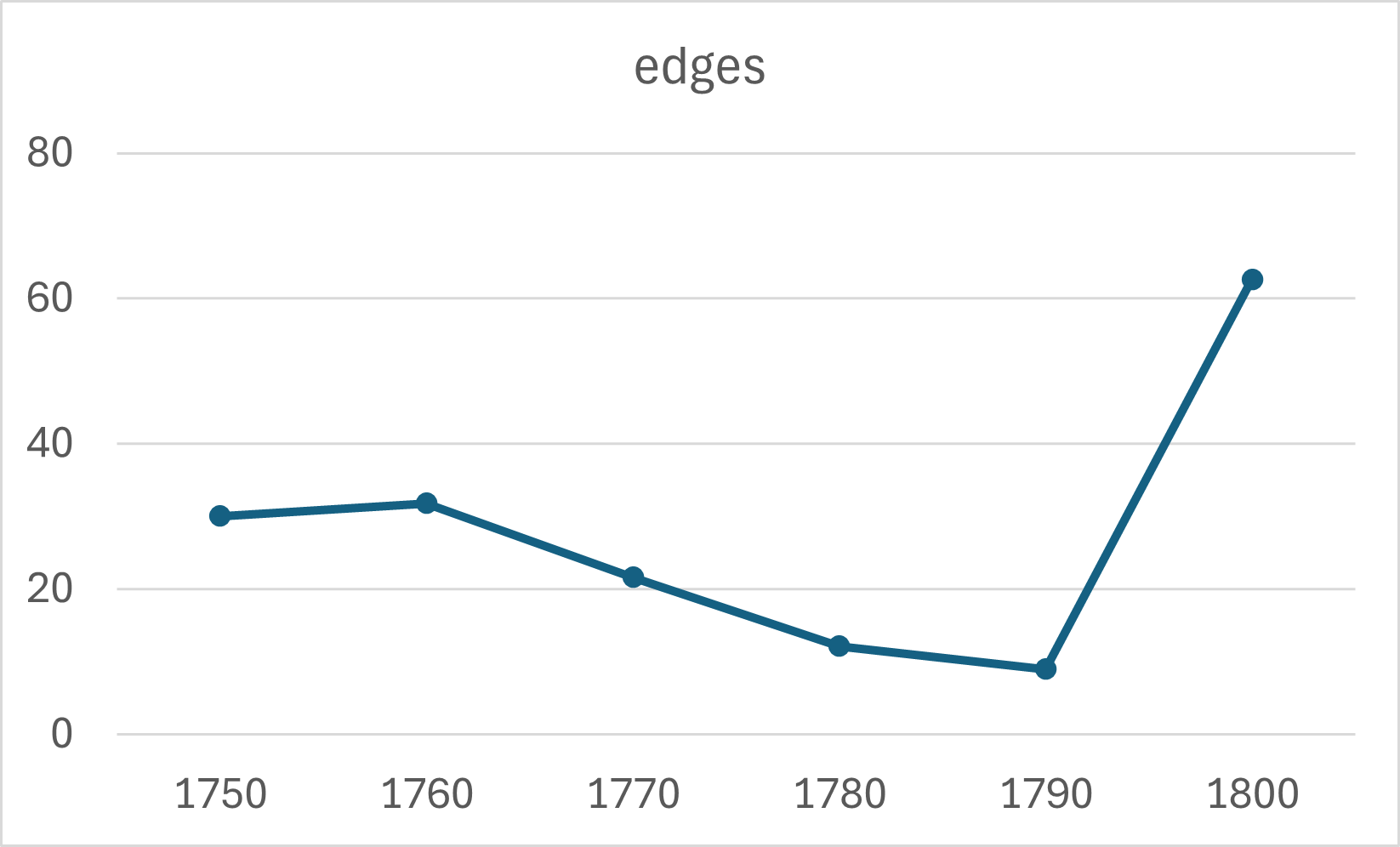} \hfill
  \includegraphics[width=0.48\linewidth]{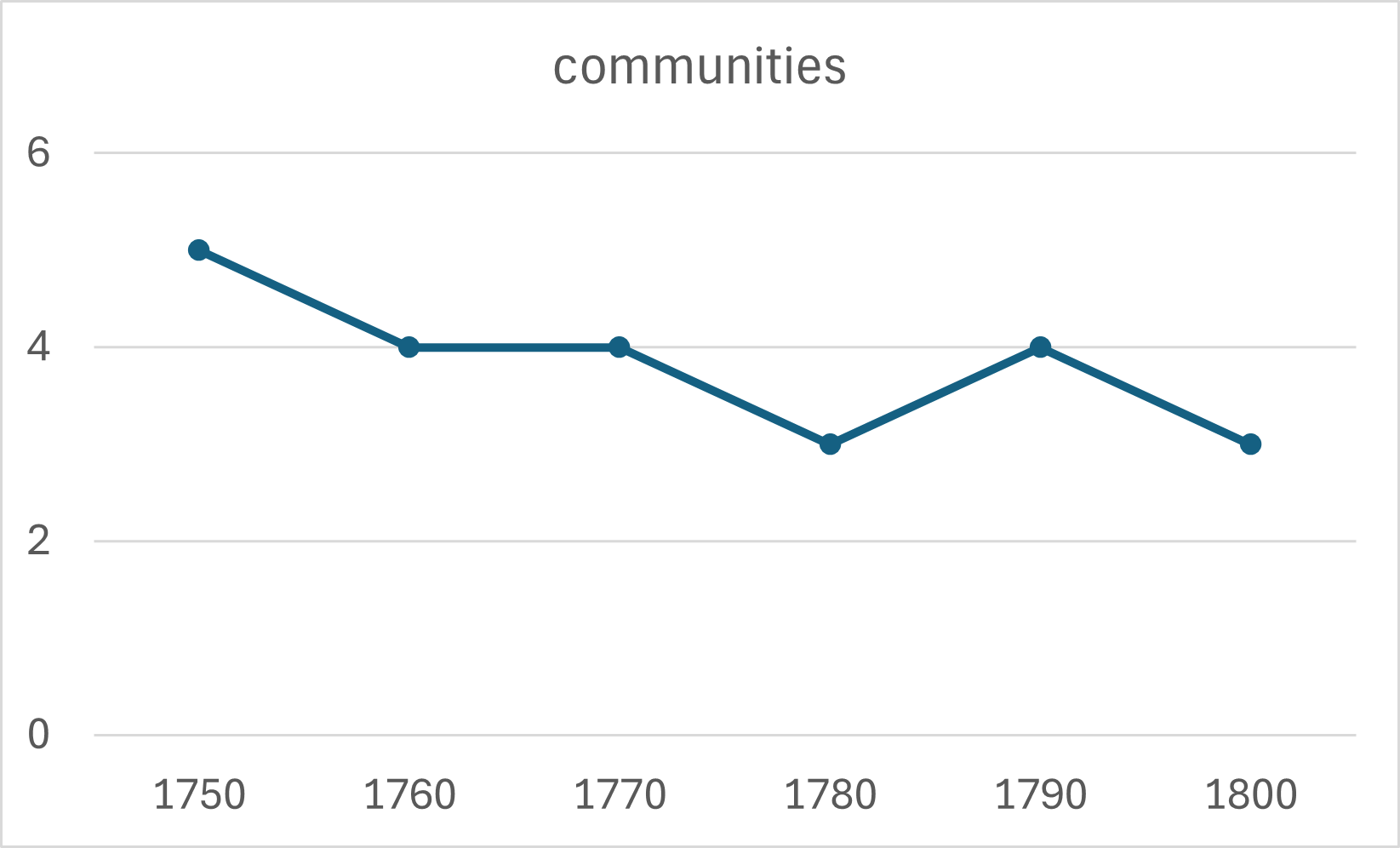} \hfill
  \includegraphics[width=0.48\linewidth]{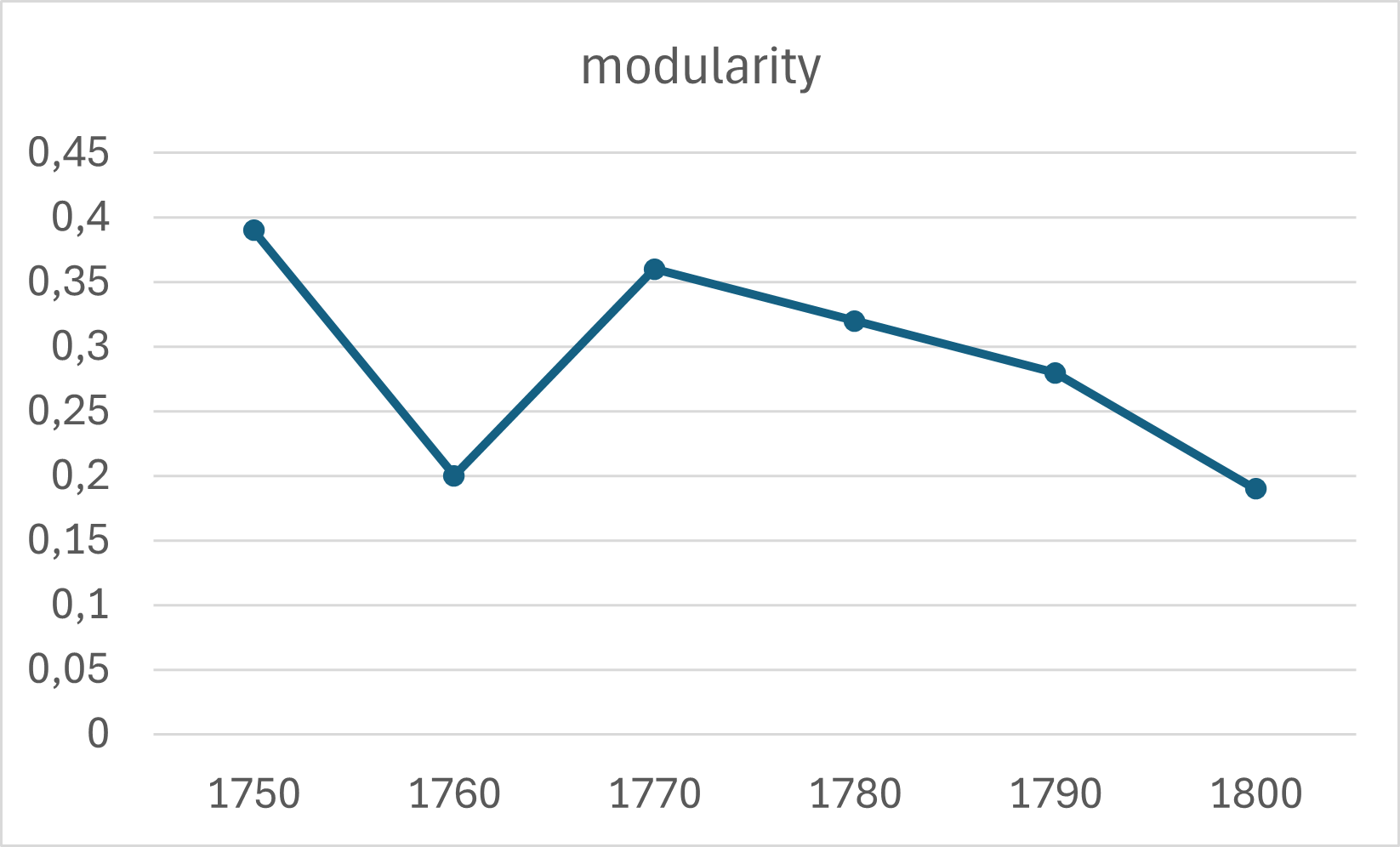} \hfill
  \includegraphics[width=0.48\linewidth]{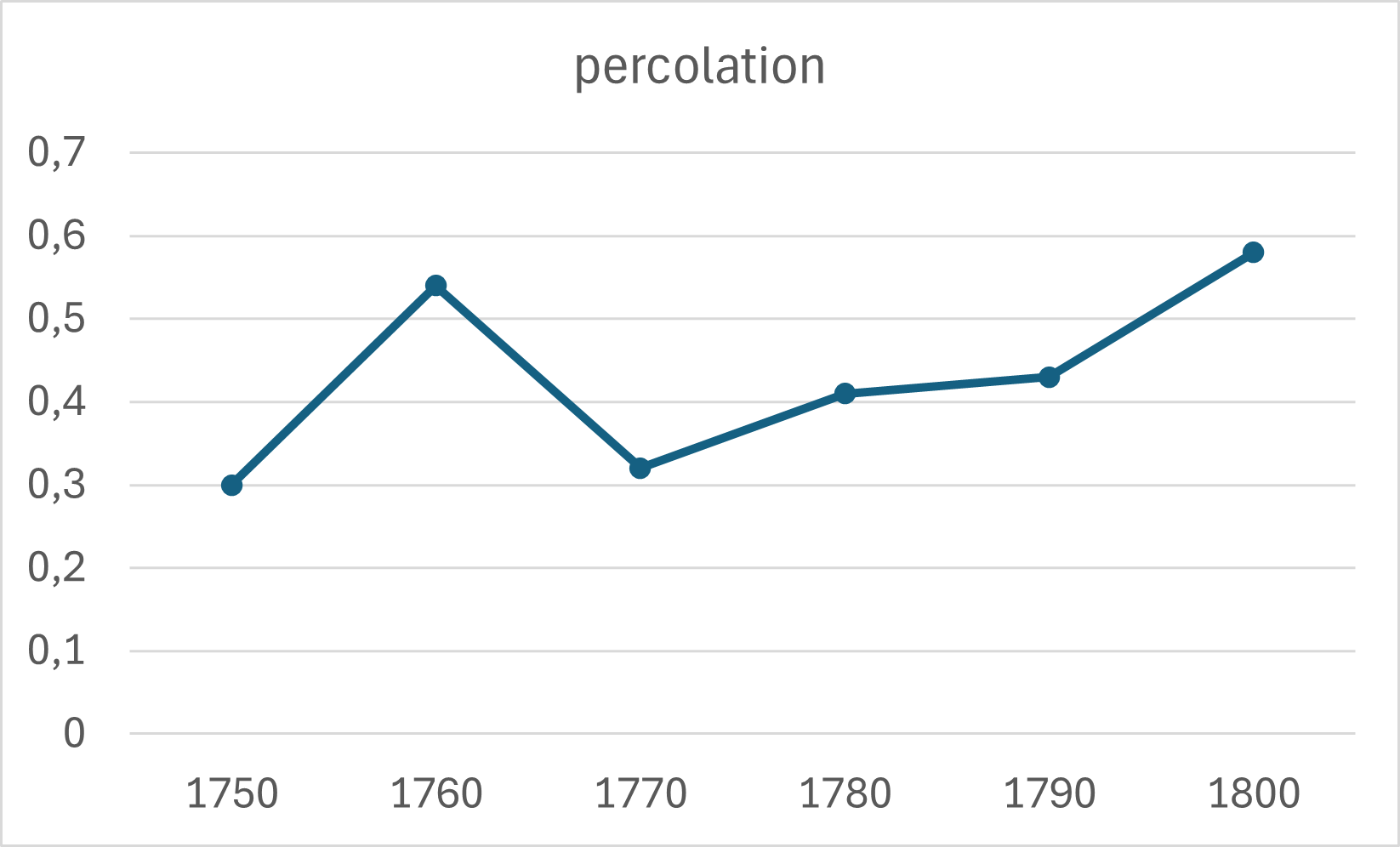} \hfill
  \caption {\textbf{Network metrics}. We used five parameters to interpret network stability over time: nodes size, edge density (where $y=1\times10^{3}$, e.g., last number of edges reported is >60 000), communities count, modularity and percolation threshold.}
  \label{app2}
\end{figure*}

\end{document}